\documentclass[sigconf]{acmart}

\renewcommand\footnotetextcopyrightpermission[1]{}
\setcopyright{none}
\usepackage{enumitem}
\usepackage{tcolorbox}
\usepackage{makecell}
\usepackage{multirow}
\AtBeginDocument{%
  }

\begin{document}



\title{Mitigating AI Risks in Computing Education via LLM-Driven Lecture Video Curation}

\author{Owen Tang}
\email{owen.tang1@unsw.edu.au}
\affiliation{%
  \institution{University of New South Wales}
  \city{Sydney}
  \state{New South Wales}
  \country{Australia}
}

\author{Alexandra Vassar}
\email{a.vassar@unsw.edu.au}
\affiliation{%
  \institution{University of New South Wales}
  \city{Sydney}
  \state{New South Wales}
  \country{Australia}
}

\author{Jake Renzella}
\email{jake.renzella@unsw.edu.au}
\affiliation{%
  \institution{University of New South Wales}
  \city{Sydney}
  \state{New South Wales}
  \country{Australia}
}

\begin{abstract}

This study evaluates the effectiveness of utilising large language models (LLMs) to retrieve targeted segments from delivered video recordings to answer student questions in introductory programming environments. By restricting AI to identifying existing, educator-verified media rather than generating open-ended text, this approach aims to mitigate common pedagogical risks such as generative hallucinations and cognitive bypassing. We benchmarked three distinct models, two proprietary (Gemini 3.1 Pro and GPT 5.4 Pro) and one open-weight (Qwen3.5 397B), against a human lecturer's manual video selections. An automated judging framework subsequently assessed the outputs for relevance, sufficiency, redundancy, and the presence of extraneous material. While the AI-retrieved timestamps rarely shared exact overlaps with the human baseline, the proprietary models achieved near-parity with the expert in delivering sufficient and highly relevant answers. Furthermore, a pilot deployment of this retrieval system in a large C programming cohort (n $\approx$ 900) demonstrated strong user engagement, with students primarily utilising the tool to review foundational concepts. By leveraging AI to retrieve established lecture material, this approach shows potential for a reliable, high-fidelity pathway for safely integrating LLMs into novice computing courses.
\end{abstract}



\maketitle

\section{Introduction}

The integration of AI into computer science education has fundamentally shifted how students engage with introductory programming. Open-domain AI (ODAI) chatbots, such as ChatGPT, have rapidly become ubiquitous tools for students navigating the complexities of CS1 \cite{Amoozadeh2024Student-AIStudents, Ghimire2024CodingCourses, Prather2023TheEducation}. However, the unrestrained reliance on these proprietary, generative systems presents significant pedagogical risks for novice learners. A key issue is model hallucination, where incorrect or logically flawed code is delivered with misleading fluency \cite{Huang2025AQuestions, Hellas2023ExploringRequests}. Furthermore, ODAI tools also suffer from complexity misalignment, where responses exceed the instructional scope and difficulty appropriate for a CS1 student \cite{Kazemitabaar2023HowEnvironment, Tie2024LLMsEngineering}. Beyond this, reliance on ODAI also poses threats to a student's cognitive development. Emerging research suggests that over-reliance on ODAI chatbots can encourage passive consumption, bypassing the methodical, high-order problem-solving practices key to early learning development \cite{Essel2024ChatGPTLLMs, Zhai2024TheReview}. Concurrently, navigating traditional course resources introduces its own friction. Although the content within multi-hour lecture recordings is still valued \cite{Case2024StudentsInstitutions}, navigating them imposes a high extraneous cognitive load \cite{Muller2024DifferencesPerspective, Chika2017LectureEducation}, on a student generation with shortening attention spans \cite{Bradbury2016AttentionMore, Crawford2025LecturesReview}.

To address this tension, we investigate a novel application of large language models (LLMs) to \textit{curate} content rather than \textit{generate}. In this study, we develop a framework to evaluate LLMs curating snippets of delivered course lecture recordings from an introductory C course in response to queries. By restricting the AI's operational knowledge base to the lecture transcripts via Retrieval-Augmented Generation (RAG), this approach preserves course-specific scope, guarantees outputs that are verified and approved for education, and minimises the risk of hallucination. 

To evaluate the feasibility and utility of this framework, this paper investigates the following research questions:
\begin{enumerate}[label=\textbf{RQ\arabic*:}, 
                  labelwidth=2.5em,   
                  labelsep=0.5em,    
                  leftmargin=!,     
                  align=left]
    \item How effective are LLMs in curating relevant snippets of lecture videos by only using their transcripts in response to a query?
    \item To what extent do the LLM curated snippets align with instructional design principles?
\end{enumerate}
Ultimately, the contributions of this work include:
\begin{enumerate}
    \item We present a novel application of LLMs to curate lecture recordings as a low-hallucination and course-aligned option for CS1 students to use AI to ask foundational queries.
    \item We demonstrate the effectiveness of proprietary models in this curation application to provide outputs that are relevant and sufficient to address novice queries.
\end{enumerate}

\section{Background and Related Work}
\subsection{Limitations and Risks of ODAI in Education}

\subsubsection{AI Hallucination}
A primary risk of utilising ODAI in educational settings is their propensity to hallucinate, where models generate false or logically flawed outputs \cite{Huang2025AQuestions, Hellas2023ExploringRequests}. Because these errors are delivered with linguistic fluency, they appear highly plausible and may mislead CS1 students who lack subject matter expertise \cite{Prather2024TheProgrammers}. Similarly, there is the risk of sycophancy, a model's tendency to prioritise user satisfaction through blind agreement rather than correcting flawed input \cite{Fanous2025SycEval:Sycophancy}. Consequently, when a novice submits a query containing core misconceptions, a sycophantic model may validate the errors rather than correct it. Because they lack the expertise to recognise and filter out instances of sycophancy, CS1 students are particularly vulnerable to this risk \cite{Bo2025InvisibleTasks}.

\subsubsection{AI Misalignment}

Even when ODAI outputs are free of hallucinations, they often fail to align with a novice's level of expertise, due to the fact that they are trained on large datasets reflecting the advanced discourse and code of experienced developers \cite{Kazemitabaar2023HowEnvironment, Tie2024LLMsEngineering}. \citeauthor{Tie2024LLMsEngineering} identified this complexity misalignment as a major failure mode for CS1 students interacting with LLMs \cite{Tie2024LLMsEngineering}. Consequently, this misalignment creates an additional cognitive burden for students to handle \cite{Tie2024LLMsEngineering, Prather2024TheProgrammers, Haindl2024DoesAnalysis}. Recent studies demonstrate that students frequently struggle to translate AI responses into their assignments \cite{Prather2024TheProgrammers}, often characterising the effort required to adapt the generated code as "high" or "very high" \cite{Haindl2024DoesAnalysis}.

\subsection{Threats to Cognitive Development}

\subsubsection{Higher Order Thinking Skills}

It is well documented that ODAI negatively impacts some cognitive behaviours of novice computer science students \cite{Xue2024DoesCS1, Zhai2024TheReview, Abuzar2025UniversityEngagement, Vaithilingam2022ExpectationModels, Vadaparty2024CS1-LLM:Instruction}. Once exposed to these dialogue agents, learners often abandon traditional educational resources in favour of faster, AI-generated solutions \cite{Xue2024DoesCS1}. This dependency shifts their approach to problem-solving, prioritising immediate outputs over methodical practice and ultimately degrading their analytical reasoning \cite{Essel2024ChatGPTLLMs, Zhai2024TheReview, Abuzar2025UniversityEngagement}. Novices typically lack the structural knowledge required to appropriately evaluate and consolidate AI-generated code \cite{Prather2024TheProgrammers, Haindl2024DoesAnalysis}. Consequently, relying on ODAI for complex problems actively hinders task-solving development \cite{Vaithilingam2022ExpectationModels} and, in some cases, can diminish students' self-confidence in their independent coding abilities \cite{Amoozadeh2024Student-AIStudents, Vadaparty2024CS1-LLM:Instruction}.

\subsubsection{Trends in Attention Spans}
Historically, student attention is claimed to decline after the first 15 minutes of a lecture \cite{Mckeachie2006MckeachieTeachers}. However, the reality today may be more concerning, as researchers assert that attention is not constant and can be influenced by many complex factors \cite{Wilson2007AttentionMinutes, Bradbury2016AttentionMore}. For example, a student's attention span is underpinned by the inherent limitations of human cognitive capacity \cite{Sweller1988CognitiveLearning}. In particular, extraneous cognitive load, which is the mental effort imposed by instructional format rather than content complexity. Because multi-hour lectures have shown to increase extraneous load \cite{Muller2024DifferencesPerspective, Chika2017LectureEducation}, it may explain why students prefer the faster responses of ODAI over traditional resources \cite{Xue2024DoesCS1}.

Despite these cognitive and delivery challenges, STEM students still heavily value lectures for their expert explanations and worked examples \cite{Case2024StudentsInstitutions}. \citeauthor{Crawford2025LecturesReview} also note a continued growth in the digitalisation and use of lecture recordings \cite{Crawford2025LecturesReview}. Therefore, innovative approaches are needed to supplement lectures. Such approaches must allow students to efficiently revisit core concepts, minimising the extraneous cognitive load from full video navigation, while actively safeguarding against the risks of ODAI.

\subsection{Lecture Content Retrieval}
\subsubsection{Lecture Chapters}
Before modern transformer-based models, automated systems like ATLAS and Content Flow Bar segmented and topically labeled lectures using machine learning trained on transition times, visual cues, and semantic annotators like Wikifier \cite{Shah2014ATLAS, PerezOrtiz2022WatchVideos}. Today, tools such as Youtube Chapters provide similar chaptering services. However, static chaptering may not be particularly beneficial to CS1 students. Because novices frequently lack the domain vocabulary to map their specific queries to broad chapter titles, they may struggle to locate relevant help. Furthermore, any thematic oversimplification inherent to broad chaptering may obscure highly relevant, granular content from the surface. Consequently, effective video navigation requires a system capable of pedagogically interpreting a student's natural language query to then determine which segments contain the solution.

\subsubsection{Temporal Video Grounding}
The task of retrieving specific video segments based on natural language queries is known as temporal sentence grounding in videos (TSGV) or video moment retrieval (VMR) and is a well researched area in the computer vision field \cite{Lin2023UniVTG:Grounding, Liu2018Cross-modalVideos, Ren2024TimeChat:Understanding}. Modern solutions typically rely on transformer-based visual feature extraction and direct semantic matching. While multimodal LLMs (MLLMs) offer advanced semantic understanding for these tasks \cite{Lu2024LLaVA-MR:Retrieval}, their success is currently constrained by the memory limitations of processing bulk video frames. Despite this, augmenting video inputs with text captions has been shown to improve MLLM performance in VMR \cite{Xie2025CaptionRetrieval}. Applying MLLMs to visually process multi-hour lecture recordings remains computationally prohibitive. However, because lecturers typically verbalise the core concepts presented on screen, text transcripts can still capture the foundational pedagogical narrative within. Therefore, utilising LLMs to retrieve segments based strictly on lecture transcripts may be an effective alternative. By leveraging their strong semantic reasoning, text-only LLMs are poised to accurately interpret both the conceptual intent behind a novice's query and the subject matter contained within the lecture.

\begin{figure*}
    \centering
    \includegraphics[width=\textwidth]{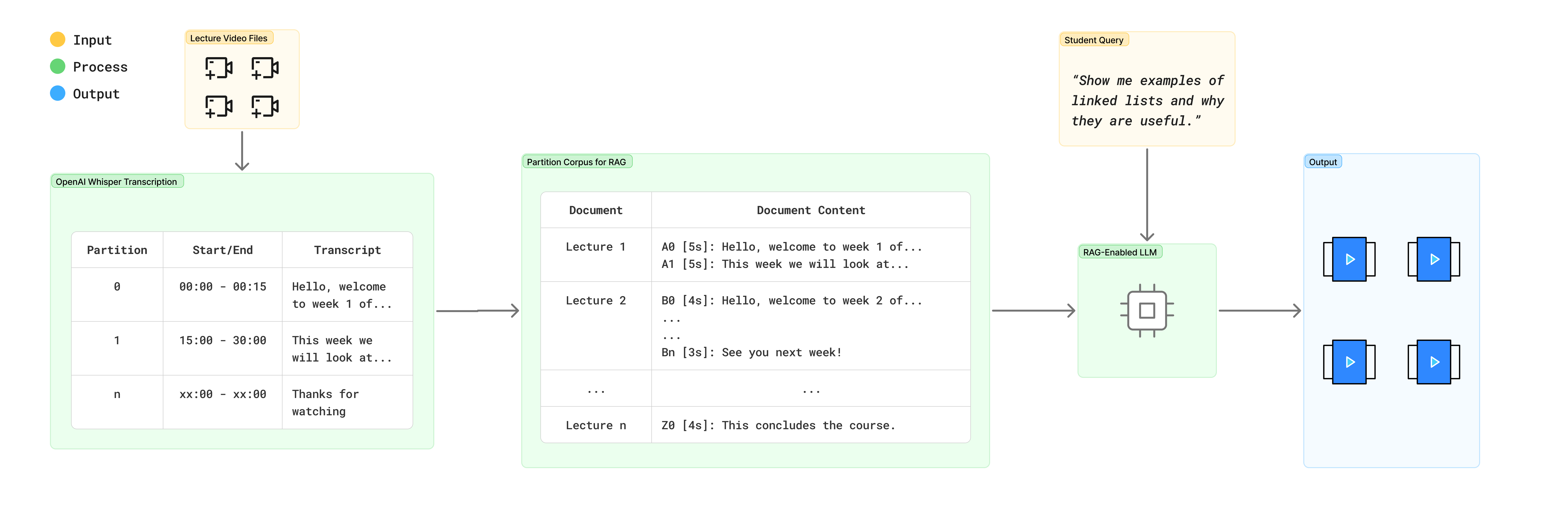}
    \caption{Pipeline of LLM Curation}  
    \Description{Pipeline of LLM Curation}
    \label{fig:pipeline}
\end{figure*}
\section{Methodology}
A custom pipeline (Figure \ref{fig:pipeline}) was built for the LLMs to curate lecture recordings from a CS1 course in C. This was implemented for each of the three distinct LLMs in this study: two proprietary models (Gemini 3.1 Pro and GPT 5.4 Pro) and one open-weight model (Qwen3.5 397B A17B). The prompt used is shown in Figure \ref{fig:prompt_template}.

\subsection{Data Pre-processing}
\subsubsection{Lecture Transcriptions}
Raw lecture videos were transcribed using OpenAI's Whisper automatic speech recognition base model. Demonstrating a low word error rate of 5\% in English contexts \cite{Radford2022RobustSupervision}, the model is highly effective at processing multilingual and multitask transcriptions \cite{Peng2023PromptingGeneralization, Amorese2023AutomaticLanguages}. This makes it suitable for accurately capturing the academic speech in this study, which predominately features stable pacing and speaker clarity.

Whisper processes the lecture recordings into short partitions containing the transcribed text, exact timestamps, and confidence metrics. To ensure transcript quality and eliminate background noise, any partition where the \texttt{no\_speech\_prob} metric exceeded a 0.6 threshold was discarded. The remaining partitions were indexed using a custom alphanumeric scheme to ensure traceability back to the original video timestamps. Lectures were assigned sequential letters and their internal partitions were numbered from zero (e.g. "B2" represents the third partition of the second lecture).

\subsubsection{LLM Configuration}
While the proprietary models manage retrieval via their native File Search\footnote{https://ai.google.dev/gemini-api/docs/file-search}\footnote{https://developers.openai.com/api/docs/guides/tools-file-search} functionalities, the open-weight Qwen model required a custom RAG architecture to ensure comparable grounding. Transcripts are partitioned into discrete chunks of 12 lines with a four-line sliding window overlap, effectively encapsulating approximately 45 seconds of spoken content. Retrieval utilises a hybrid ensemble that weights BM25 (sparse keyword search) and FAISS (dense semantic vector search) at a ratio of 0.85 to 0.15, respectively. Semantic embeddings for the FAISS index were generated using the qwen3-embedding-8b model. The retriever is configured to extract the top ten most relevant chunks ($k=10$). Alternative configurations, such as relying exclusively on dense retrieval, increasing chunk sizes, or reducing retrieval depth, caused significant performance degradation, frequently resulting in empty outputs or the retrieval of irrelevant background noise.

\begin{figure}[H]
    \centering
    \begin{tcolorbox}[colback=gray!5, colframe=black, fonttitle=\bfseries, boxrule=0.5pt, arc=2mm, left=3mm, right=1mm, top=0.5mm, bottom=0.5mm,before skip=0pt, after skip=0pt]

    \footnotesize
    
    \textit{You have access to lecture transcripts divided into sequential sections. Each section is labelled with an alphanumeric index (e.g., A0, B1) followed by its duration in seconds. Your task is to return only the indices of sections that most directly address the student's question. Do not include irrelevant or introductory content — begin from where the concept is taught in detail. Only return a set of indices (e.g., (A0-A7, C9-C11) or ()). If the question is irrelevant to all transcripts, return ().\\
    Student Question: [\textbf{QUERY}]. Please identify the most relevant transcript section indices that best answer this question.}
    \end{tcolorbox}
    \caption{Prompt Template}
    \Description{Prompt Template}
    \label{fig:prompt_template}
\end{figure}


\section{Evaluation}
The models are evaluated on a set of queries against a human expert baseline, the course lecturer. Their outputs are assessed in two phases: a quantitative analysis measuring the temporal overlap and boundary similarity between the LLMs and the lecturer, and a blind, LLM-as-a-judge qualitative evaluation of the outputs.

\subsection{Query Set}\label{sec:query_set}
To represent the diverse cognitive demands expected in a CS1 course, a set of five test queries (Table \ref{tab:query_set}) was developed using Bloom's revised taxonomy (BRT) \cite{Krathwohl2002AOverview}. BRT classifies learning objectives across two dimensions: cognitive processes (Remember, Understand, Apply, Analyse, Evaluate, and Create) and knowledge types (Factual, Conceptual, Procedural, and Metacognitive). Recent studies have utilised BRT to map student behaviors in digital pedagogical tools onto cognitive learning dimensions \cite{Wedlock2017TheGenerations, Vavilina2020USINGCLASSROOM} and demonstrate how these tools facilitate both low and high-order thinking \cite{Alaghbary2021IntegratingLearning}.

Intersecting these two dimensions provides a structured foundation to systematically assess the LLMs' curation capabilities across various learning stages. It is noted that the final stages of both dimensions, Create and Metacognitive, were excluded as they usually involve extended planning and self-reflection that fall outside the scope of querying a lecture recording. To prevent evaluation overlap, we restricted specific intersections. Remember is paired exclusively with Factual knowledge to establish a baseline for retrieving discrete syntax. Understand and Analyse are mapped to Conceptual knowledge to evaluate the models' ability to locate theoretical reasoning. Finally, Apply and Evaluate are aligned with Procedural knowledge to examine the retrieval of practical code demonstrations. This targeted selection ensures each query assesses a strictly distinct, non-overlapping learning process.

\begin{table}[H]
    \caption{Query Set used in Prompts}
    \label{tab:query_set}
  \begin{tabular}{cp{3.75cm}cc}
    \toprule
    \textbf{ID} & \textbf{Query} & \textbf{BRT CP} & \textbf{BRT KT} \\
        \midrule
        Q1 & "What is the syntax for declaring a pointer in C?" & Remember& Factual\\
        Q2 & "Explain how a linked list differs from an array in terms of memory allocation." & Understand & Conceptual\\
        Q3 & "How do I write the code to insert a node at the beginning of a linked list?" & Apply& Procedural\\
        Q4 & "Why is it more efficient to use a linked list instead of an array for frequent data insertions?" & Analyse& Conceptual\\
        Q5 & "I have a memory leak in my C program, how can I identify and resolve it?" & Evaluate& Procedural\\
  \bottomrule
\end{tabular}
\end{table}
\subsection{Baseline Comparisons}\label{sec:evaluation_baseline}
One method of evaluating the LLMs' selection of snippets is to compare it with what the course lecturer would have recommended to students if asked the same query. These selected time frames are quantitatively compared for similarity to those selected by the LLMs. Given the similarity of this task to video moment retrieval, the metrics used are widely recognised as a standard in that field \cite{Lin2023UniVTG:Grounding, Lu2024LLaVA-MR:Retrieval, Xie2025CaptionRetrieval, Ren2024TimeChat:Understanding, Gao2017TALL:Query}.

To measure the temporal overlap between the lecturer-selected time frames (H) and the LLM-selected time frames (A), we calculate the intersection over union (IoU) for each response. IoU provides a direct indication of how precisely the LLM-curated time frames mirror the selections of the lecturer. To complement the overall alignment measured by IoU, precision (P) and recall (R) are utilised to provide deeper insight into the curation quality. Precision measures the proportion of the LLM's selections that fall within the lecturer's time frames, thereby penalising the inclusion of other content. Recall measures the coverage of the lecturer's selections captured by the LLM. Together, these metrics can reveal whether the LLM's curation, in relation to the lecturer, is overly verbose (low precision) or structurally deficient (low recall). The formulas for IoU, P, and R are defined as follows:

\begin{displaymath}
\textbf{IoU} = \frac{|\textbf{H} \cap \textbf{A}|}{|\textbf{H} \cup \textbf{A}|}, \quad \textbf{P} = \frac{|\textbf{H} \cap \textbf{A}|}{|\textbf{H}|}, \quad \textbf{R} = \frac{|\textbf{H} \cap \textbf{A}|}{|\textbf{A}|}
\end{displaymath}

\subsection{LLM-as-a-Judge}\label{sec:evaluation_llm_judge}
To evaluate the utility of the retrieved segments beyond mere temporal agreement, we employ the LLM-as-a-judge methodology, an approach widely used for assessing pedagogical feedback \cite{LeeSolano2026Fine-TuningMessages, Koutcheme2024OpenGPT-4-As-A-Judge, Koutcheme2025EvaluatingFeedback}. Our evaluation utilises a diverse ensemble of judges comprising both proprietary (Claude Sonnet 4.6 and Gemini 3.1 Pro) and open-weight (Mistral Large) models. Because including Gemini as a judge risks introducing self-preference bias toward its own outputs, we enforce an agreement rule across the ensemble. Recent studies on ensemble judges have shown similar strategies can mitigate this bias and still yield robust evaluations \cite{Seo2025LargeAccuracy, Verga2024ReplacingModels}.

Each judge independently assesses the transcript outputs using the scoring rubric detailed in Table \ref{tab:rubric_criteria}. This rubric evaluates performance across two primary dimensions: ($C_{1-2}$) and instructional design ($C_{3-4}$), each on a 5-point Likert scale (1 = Strongly Disagree to 5 = Strongly Agree), including a worded justification for each score. To ensure rigorous consensus, a final average score is only accepted if the range of scores across all judges is at most 1. Specifically, for a given criterion within a given query, the agreement rule is defined as follows:
\begin{displaymath}
\max(x_{Cla}, x_{Gem}, x_{Mis}) - \min(x_{Cla}, x_{Gem}, x_{Mis}) <= 1
\end{displaymath}
\subsection{Evaluation Rubric} 
\begin{table}[h]
    \caption{Evaluation Rubric}
    \label{tab:rubric_criteria}
    \begin{tabular}{cp{6.6cm}}
        \toprule
        \textbf{ID} & \textbf{Criterion}\\
        \midrule
        $\textbf{C}_\textbf{1}$ & "The lecture content in the snippet(s) is relevant to the query."\\
        $\textbf{C}_\textbf{2}$ & "The lecture content in the snippet(s) is sufficient to address the query."\\
        $\textbf{C}_\textbf{3}$ & "The lecture content of each snippet does not repeat information across the returned snippets."\\
        $\textbf{C}_\textbf{4}$ & "The snippet(s) do not contain irrelevant content (e.g. conversational tangents, prefaces, crowd control)."\\
        \bottomrule
    \end{tabular}
\end{table}
\begin{table*}
\centering
\caption{IoU, P, R, and Criterion Scores (Agreed) Across 5 Queries}
\label{tab:results}
\begin{tabular}{c | ll | ll | ll | ll | ll| ll}
\toprule
\textbf{Model} & 
\multicolumn{2}{c|}{\textbf{Q1}} & 
\multicolumn{2}{c|}{\textbf{Q2}} & 
\multicolumn{2}{c|}{\textbf{Q3}} & 
\multicolumn{2}{c|}{\textbf{Q4}} & 
\multicolumn{2}{c|}{\textbf{Q5}} & 
\multicolumn{2}{c}{\textbf{Averages}\footnotemark} \\
\midrule
\multirow{4}{*}{\makecell{\textbf{Gemini} \\ \textbf{3.5 Pro} \\ ($P_A=\frac{18}{20}=90\%$)}} 
& IoU: 0.45 & $C_1$: 5.0 & IoU: 0.18 & $C_1$: 5.0 & \multicolumn{2}{c|}{$C_1$: 5.0} & IoU: 0.0 & $C_1$: 5.0 & \multicolumn{2}{c|}{$C_1$: 4.33} & \textbf{IoU: 0.21} & $\textbf{C}_\textbf{1}$\textbf{: 4.87} \\
& P: 0.5 & $C_2$: 5.0 & P: 0.18 & $C_2$: 5.0 & \multicolumn{2}{c|}{$C_2$: 4.33} & P: 0.0 & $C_2$: 4.33 & \multicolumn{2}{c|}{$C_2$: 2.33} & \textbf{P: 0.23} & $\textbf{C}_\textbf{2}$\textbf{: 4.2} \\
& R: 0.8 & $C_3$: 2.33 & R: 1.0 & $C_3$: None & \multicolumn{2}{c|}{$C_3$: None} & R: 0.0 & $C_3$: 2.0 & \multicolumn{2}{c|}{$C_3$: 2.0} & \textbf{R: 0.6} & $\textbf{C}_\textbf{3}$\textbf{: 2.11} \\
&         & $C_4$: 2.0 &         & $C_4$: 1.67 & \multicolumn{2}{c|}{$C_4$: 2.0} &         & $C_4$: 2.33 & \multicolumn{2}{c|}{$C_4$: 2.0} &         & $\textbf{C}_\textbf{4}$\textbf{: 2.0} \\
\midrule
\multirow{4}{*}{\makecell{\textbf{GPT} \\ \textbf{5.4 Pro} \\ ($P_A=\frac{16}{20}=80\%$)}} 
& IoU: 0.34 & $C_1$: 5.0 & IoU: 0.32 & $C_1$: 5.0 & \multicolumn{2}{c|}{$C_1$: 5.0} & IoU: 0.26 & $C_1$: 5.0 & \multicolumn{2}{c|}{$C_1$: 4.33} & \textbf{IoU: 0.31} & $\textbf{C}_\textbf{1}$\textbf{: 4.87} \\
& P: 1.0 & $C_2$: 4.67 & P: 0.32 & $C_2$: 4.67 & \multicolumn{2}{c|}{$C_2$: 4.33} & P: 0.26 & $C_2$: 4.33 & \multicolumn{2}{c|}{$C_2$: 2.33} & \textbf{P: 0.53} & $\textbf{C}_\textbf{2}$\textbf{: 4.07} \\
& R: 0.34 & $C_3$: 5.0 & R: 1.0 & $C_3$: None & \multicolumn{2}{c|}{$C_3$: None} & R: 1.0 & $C_3$: None & \multicolumn{2}{c|}{$C_3$: None} & \textbf{R: 0.78} & $\textbf{C}_\textbf{3}$\textbf{: 5.0} \\
&         & $C_4$: 4.0 &         & $C_4$: 2.0 & \multicolumn{2}{c|}{$C_4$: 2.0} &         & $C_4$: 3.0 & \multicolumn{2}{c|}{$C_4$: 2.0} &         & $\textbf{C}_\textbf{4}$\textbf{: 2.6} \\
\midrule
\multirow{4}{*}{\makecell{\textbf{Qwen3.5} \\ \textbf{397B} \\ ($P_A=\frac{14}{20}=70\%$)}} 
& IoU: 0.0 & $C_1$: 4.67 & IoU: 0.16 & $C_1$: 5.0 & \multicolumn{2}{c|}{$C_1$: None} & IoU: 0.0 & $C_1$: 2.0 & \multicolumn{2}{c|}{$C_1$: None} & \textbf{IoU: 0.05} & $\textbf{C}_\textbf{1}$\textbf{: 3.89} \\
& P: 0.0 & $C_2$: 3.0 & P: 0.23 & $C_2$: None & \multicolumn{2}{c|}{$C_2$: 1.33} & P: 0.0 & $C_2$: 1.0 & \multicolumn{2}{c|}{$C_2$: 2.33} & \textbf{P: 0.08} & $\textbf{C}_\textbf{2}$\textbf{: 1.92} \\
& R: 0.0 & $C_3$: 5.0 & R: 0.33 & $C_3$: None & \multicolumn{2}{c|}{$C_3$: None} & R: 0.0 & $C_3$: 4.67 & \multicolumn{2}{c|}{$C_3$: 5.0} & \textbf{R: 0.11} & $\textbf{C}_\textbf{3}$\textbf{: 4.89} \\
&         & $C_4$: 2.33 &         & $C_4$: 4.33 & \multicolumn{2}{c|}{$C_4$: 2.0} &         & $C_4$: 1.67 & \multicolumn{2}{c|}{$C_4$: None} &         & $\textbf{C}_\textbf{4}$\textbf{: 2.58} \\
\midrule
\multirow{4}{*}{\makecell{\textbf{Lecturer} \\ ($P_A=\frac{10}{12}=83\%$)}} 
& \multicolumn{2}{c|}{$C_1$: 5.0} & \multicolumn{2}{c|}{$C_1$: 5.0} & \multicolumn{2}{c|}{-} & \multicolumn{2}{c|}{$C_1$: 4.67} & \multicolumn{2}{c|}{-} & \multicolumn{2}{c}{$\textbf{C}_\textbf{1}$\textbf{: 4.89}} \\
& \multicolumn{2}{c|}{$C_2$: 4.33} & \multicolumn{2}{c|}{$C_2$: 3.67} & \multicolumn{2}{c|}{-} & \multicolumn{2}{c|}{$C_2$: None} & \multicolumn{2}{c|}{-} & \multicolumn{2}{c}{$\textbf{C}_\textbf{2}$\textbf{: 4.0}} \\
& \multicolumn{2}{c|}{$C_3$: None} & \multicolumn{2}{c|}{$C_3$: 5.0} & \multicolumn{2}{c|}{-} & \multicolumn{2}{c|}{$C_3$: 5.0} & \multicolumn{2}{c|}{-} & \multicolumn{2}{c}{$\textbf{C}_\textbf{3}$\textbf{: 5.0}} \\
& \multicolumn{2}{c|}{$C_4$: 2.33} & \multicolumn{2}{c|}{$C_4$: 1.67} & \multicolumn{2}{c|}{-} & \multicolumn{2}{c|}{$C_4$: 3.33} & \multicolumn{2}{c|}{-} & \multicolumn{2}{c}{$\textbf{C}_\textbf{4}$\textbf{: 2.44}} \\
\bottomrule
\end{tabular}
\end{table*}
To address RQ1, we measure the relevance of the extracted snippets to the queries. Because this task mirrors traditional Information Retrieval (IR) tasks, we ground this section of the evaluation in \citeauthor{Saracevic2007Relevance:Relevance}'s model of relevance \cite{Saracevic2007Relevance:Relevance}. \citeauthor{Saracevic2007Relevance:Relevance} posits that relevance is not a binary metric but spans multiple dimensions: algorithmic, topical, cognitive, situational, and affective. The importance of these dimensions varies by domain, with fields like e-commerce and law prioritising them differently \cite{vanOpijnen2017OnRetrieval, Tsagkias2021ChallengesRecommendations}. In the context of CS1 education and RQ1, the aim is to ensure the retrieved content matches the topic within the query and that the query is sufficiently addressed. Therefore, $C_1$ evaluates topical relevance by measuring whether the snippet matches the query's core subject. $C_2$ evaluates situational relevance, ensuring the output has sufficient coverage to completely address the query.

To address RQ2, the structural quality of the outputs is evaluated, inspired by \citeauthor{Sweller1988CognitiveLearning}'s cognitive load theory (CLT) \cite{Sweller1988CognitiveLearning} and \citeauthor{Mayer2021MultimediaLearning}'s principles for multimedia instructional design \cite{Mayer2021MultimediaLearning}. CLT warns against the redundancy effect, where duplicate information consumes limited working memory. This forms $C_3$, which penalises redundant explanations across the returned snippets. \citeauthor{Mayer2021MultimediaLearning}'s coherence principle dictates that learning improves significantly when extraneous media and words are excluded \cite{Mayer2021MultimediaLearning}. $C_4$ aims to measure how well the models avoided including such irrelevant content such as conversational tangents, prefaces, and crowd control.

\section{Pilot Deployment}
The curation tool, using Gemini 2.5 Pro, was deployed as a supplementary resource in a ten-week introductory C course ($n=903$ students). Students gained unrestricted access in week three, with the transcript corpus continuously updated after each twice-weekly lecture. Student usage data was tracked in every interaction to gain insights into their usage of tool. This included their queries, the returned snippets, viewership behaviours, and voluntary user ratings approving or disapproving snippet relevance.

\label{sec:discussion}
\section{Results \& Discussion}
\subsection{Temporal Comparison and Rubric Evaluation}
Table \ref{tab:results} presents temporal overlap metrics (IoU, P, R) and qualitative judgements ($C_{1-4}$) for all queries and curators. The proportion of judge agreement is denoted by $P_A$, and  "None" if the agreement rule was not satisfied. Note that the lecturer did not produce any response for Q3 and Q5 as they "did not find anything to directly address the queries completely". Accordingly, there are no temporal overlap metrics for those queries.
\footnotetext{The averages for $C_{1-4}$ are calculated on agreed values only.}

\subsubsection{Proprietary Models}
The temporal overlap metrics show a difference in curation strategies between human educators and LLMs. The proprietary models generally exhibited high recall (GPT: 0.78, Gemini: 0.60) but much lower precision and IoU (GPT: $P=0.53$, $IoU=0.31$, Gemini: $P=0.23$, $IoU=0.21$). This discrepancy indicates that while these LLMs captured much of the core material targeted by the lecturer, they cast a wider net that included more explanations. Despite this, these models were found to still produce highly applicable outputs. Overall, both proprietary models achieved near-perfect scores for $C_1$, with judges noting that outputs "directly address the core" of the queries by surfacing "highly relevant" fundamental concepts. They also showed a strong ability to provide sufficient information ($C_2$), where outputs were described as "abundantly sufficient" and capable of "fully resolving the student's question" across a diverse range of introductory topics.

For Q3 and Q5, both models attempted to fulfill the queries, despite the lecturer not being able to do so. For Q3, the models outputs were found to be both highly relevant and sufficient ($C_1=5.0$, $C_2=4.33$). For Q5, the models still returned content that was highly relevant ($C_1=4.33$) but ultimately insufficient ($C_2=2.33$). The LLM judges noted in their justifications that the Q5 segments lacked the explicitly requested techniques for testing and treating memory leaks, instead focusing on the related concepts of malloc() and free(). Nevertheless, the proprietary models showed that they would still surface highly relevant concepts even when the direct answer is absent, and that in some cases (Q3), it was sufficient to address the query. This is a notable contrast in how proprietary models and the lecturer handle content gaps. While the lecturer was reluctant to choose a snippet that did not explicitly answer the query, the proprietary LLMs insisted on attempting to deliver the most relevant information available. This behaviour can be highly valuable for pointing students in the right direction, however it is crucial that it maintains high relevance to avoid misleading them.

Evaluating $C_3$ and $C_4$ reveals challenges in achieving optimal conciseness across both automated models and the human baseline. The proprietary LLMs exhibited mixed performance on these metrics and the majority of disagreement in scoring. Interestingly, the human lecturer also demonstrated a poor result in $C_4$ (2.44), frequently noted by the LLM judges to contain "tangential discussions". This may suggest that the task of cleanly segmenting spoken lecture transcripts without sacrificing the narrative flow required for effective student comprehension may be inherently difficult.

\subsubsection{Open Source Model}
Qwen3.5 struggled with much of this curation task compared to its proprietary counterparts. By yielding near-zero averages for IoU (0.05), precision (0.08), and recall (0.11), its selections were completely dissimilar to the lecturer's. However, unlike the proprietary models, the low temporal matching did eventuate in poor sufficiency ($C_2=1.92$) and fair relevance ($C_1=3.89$). Its outputs were described as sometimes "directly related" but frequently "insufficient" and "lacking detail". Conversely, it performed well in reducing redundant information ($C_3=4.89$), a criterion that the other models had either poor or disagreed performance. Ultimately, open-source models offer distinct advantages regarding data privacy and cost, and Qwen's performance demonstrates a potential to improve, rather than a complete failure. However, to match the high relevancy and sufficiency seen from its proprietary counterparts, further work is needed in optimising model selection and parameter configuration.

\subsection{Pilot Deployment}
\subsubsection{Student Engagement}
The pilot deployment showed an early optimism for the lecture curation tool. Over seven weeks, the tool processed 343 prompts and maintained a 58\% return-user rate. The delivered snippets were short and had a median duration of just 91 seconds. Student feedback, although limited, strongly approved of the outputs, with an overwhelmingly positive ratio of 19 approvals to 2 disapprovals. This aligns with the strong relevance ($C_1$) and sufficiency ($C_2$) scores for Gemini in the LLM-as-a-judge evaluation. Notably, sessions with positive feedback had a significantly shorter average watch time (46 seconds) compared to the overall average (126 seconds), suggesting that when the tool works successfully, it allows students to find exactly what they need in a short time. 
\subsubsection{Query Trends}
An analysis of query trends demonstrates that students primarily utilised the tool to clarify fundamental C programming concepts. The most frequently queried keywords aligned directly with the core curriculum, heavily featuring arrays (48 queries), linked lists (45 queries), and pointers (40 queries). Furthermore, query intent was highly focused on consolidating fundamental concepts as nearly half (42.9\%) of all submissions were direct search terms designed to retrieve specific lecture excerpts, while the remainder were split between definitional ("What is...") and procedural ("How to...") questions. This clustering reinforces the idea that CS1 students most often seek help on fundamental topics and that AI tools should align with these needs by producing reliable content and maintaining course-specific fidelity.

\label{sec:lim_future_work}
\section{Limitations and Future Work}
Several methodological limitations motivate future research from this study. Firstly, the quality of the LLM curated output will always inherently depend on the raw lecture material and transcription accuracy. Although our evaluation utilised an LLM-as-a-judge configured with independent criteria, the transcription of Whisper is not perfect. Secondly, the transcript-based evaluation could not account for visual cues like on-screen code demonstrations. Even if the impact is limited, as demonstrations are typically also verbalised, future work will involve human experts evaluating the actual video snippets to validate these findings. Thirdly, the human baseline was restricted to a single lecturer to ensure accurate alignment with the specific nuances of their course. Future iterations could broaden this baseline by expanding to multiple courses with shared curricula and also include more diverse queries. Furthermore, because our pilot deployment focused strictly on observing general engagement with the tool, we did not track demographic context. Subsequent longitudinal studies should do so to empirically measure the tool's impact on student learning outcomes and learning gains. Due to the high number of unrestricted users and budgetary constraints, the tool used an earlier version of Gemini Pro, which may differ slightly in performance. Lastly, while Qwen3.5 exhibited low baseline performance under its current configuration, optimising open-source configurations remains a key future work given its benefits. 

\section{Conclusion}
This study demonstrates that utilising LLMs for lecture curation offers a viable, course-aligned option for using AI in CS1. To address RQ1, we found that the proprietary models curated highly relevant and sufficient content despite having low temporal overlap with the lecturer's choices. They managed this even in instances where the lecturer found no suitable material. This is an interesting finding, as it shows an intriguing behavioural divergence when a direct answer is absent. While the lecturer refrained from providing a response that falls short of directly addressing the query, the proprietary LLMs persistently attempted to bridge the gap by surfacing the closest foundational concepts. Regarding RQ2, the LLMs' outputs exhibited mixed performances in avoiding redundant explanations but were found to perform just as poorly as the human lecturer in avoiding irrelevant conversational tangents. Ultimately, alongside early signs of positive student reception in a pilot deployment, this framework has strong potential in minimising the pedagogical risks of ODAI such as AI hallucination and complexity misalignment. In an educational landscape increasingly dominated by ODAI, embedding these safeguards is paramount to designing educational tools that support, rather than hinder, foundational learners.

\bibliographystyle{ACM-Reference-Format}
\bibliography{references}

\end{document}